\documentclass[submitting]{nst}

\usepackage{lineno}

\usepackage[utf8]{inputenc}
\usepackage{subfigure,dcolumn}
\usepackage[T2A,T1]{fontenc}
\usepackage[russian,english]{babel}
\usepackage{multirow}
\usepackage{booktabs}
\usepackage{listings}
\usepackage{color}
\usepackage{graphicx} 
\usepackage{epstopdf}

\usepackage{natbib}
\let\footnote\relax

\let\textcite\relax

\let\citeauthor\relax
\let\citeyear\relax
\expandafter\let\csname
ver@natbib.sty\endcsname\relax

\usepackage[
    backend=biber,
    bibencoding=utf8,
    style=numeric-comp,
    giveninits=true,
    autolang=other, 
    sorting=none,
    doi=true, isbn=false, eprint=false, url=false, date=year, 
    maxcitenames=3,
    mincitenames=3,
    maxbibnames=3,
    minbibnames=3,
]{biblatex}

\usepackage{csquotes}
\renewbibmacro{in:}{}
\begin{document}
\renewcommand{\bibliography}[1]{}

\title{Method for detector description transformation to Unity and application in BESIII}
\author{Kai-Xuan Huang}
\affiliation{School of Physics, Sun Yat-sen University, Guangzhou 510275, China}
\author{Zhi-Jun Li}
\affiliation{School of Physics, Sun Yat-sen University, Guangzhou 510275, China}
\author{Zhen Qian}
\affiliation{School of Physics, Sun Yat-sen University, Guangzhou 510275, China}
\author{Jiang Zhu}
\affiliation{School of Physics, Sun Yat-sen University, Guangzhou 510275, China}
\author{Hao-Yuan Li}
\affiliation{School of Physics, Sun Yat-sen University, Guangzhou 510275, China}
\author{Yu-Mei Zhang}
\email[Yu-Mei Zhang, ]{zhangym26@mail.sysu.edu.cn}
\affiliation{Sino-French Institute of Nuclear Engineering and Technology, Sun Yat-sen University, Zhuhai 519082, China}
\author{Sheng-Sen Sun}
\affiliation{Institute of High Energy Physics, Chinese Academy of Sciences, Beijing 100049, China}
\affiliation{University of Chinese Academy of Sciences, Beijing, 100049, China}
\author{Zheng-Yun You}
\email[Zheng-Yun You, ]{youzhy5@mail.sysu.edu.cn}
\affiliation{School of Physics, Sun Yat-sen University, Guangzhou 510275, China}

\begin{abstract}

Detector and event visualization are essential parts of the software used in high-energy physics (HEP)  experiments.
Modern visualization techniques and multimedia production platforms such as Unity provide impressive display effects and professional extensions for visualization in HEP experiments.
In this study, a method for automatic detector description transformation is presented, which can convert the complicated HEP detector geometry from GDML in offline software to 3D modeling in Unity.
The method was successfully applied in the BESIII experiment and can be further developed into applications such as event displays , data monitoring, or virtual reality.
It has great potential in detector design, offline software development, physics analysis, and outreach for next-generation HEP experiments as well as applications in nuclear techniques for the industry.

\end{abstract}
 
\keywords{Detector description, Visualization, Unity, GDML, BESIII}

\maketitle
\nolinenumbers
\section{Introduction}

Detector description and visualization play important roles in various aspects of the life cycle of a high-energy physics (HEP) experiment, including detector design, optimization, simulation, reconstruction, detector commissioning, monitoring, event display, physics analysis, outreach, and education.
In the HEP Software Foundation (HSF) Community White Paper~\cite{HSF} and the Roadmap for HEP Software and Computing R\&D for the 2020s~\cite{Roadmap}, suggestions and guidelines for visualization tools and techniques in future experiments have been specifically discussed, focusing on detector geometry visualization, event display, and interactivity.

The detectors in HEP experiments are usually large-scale scientific apparatuses with complicated geometries, which are composed of millions of detector units, such as the ATLAS~\cite{atlas} and CMS~\cite{CMS} detectors at the Large Hadron Collider (LHC)~\cite{LHC}. The detector description is developed based on professional geometry toolkits, such as the commonly used Geometry Description Markup Language (GDML)~\cite{GDML} and Detector Description Toolkit for High Energy Physics (DD4hep)~\cite{dd4hep}. 
Detector geometry information in GDML or DD4hep format can be imported into the offline software of an experiment to provide a consistent detector description for different applications~\cite{BESIII_geo,ROOT_Geant4}.  

However, displaying complicated detectors in offline software is difficult because the HEP software community does not have a common visualization tool that meets the requirements of visualization effects, speed, efficiency, and portability.
The ROOT software~\cite{ROOT} and its EVE package~\cite{ROOT_EVE} have been used in some experiments for detector visualization and event display, such as ALICE~\cite{ALICE}, BESIII~\cite{BESIII}, and JUNO~\cite{JUNO_eventdisplay,JUNO_liquid}. 
Although ROOT-based event display tools have the advantage of their development in offline software frameworks, their visualization effects are limited, and they are also difficult to implement on platforms other than Linux.  

In recent years, popular video game engines from the industry, such as Unity~\cite{Unity}, have been used in HEP experiments for detector visualization and event display. 
Unity is a cross-platform engine for creating  games, architecture, videos, and animation. 
It supports more than 20 different platforms, including Windows, Linux, macOS, iOS, and Android, which makes Unity the most popular application development platform on Apple Store and Google Play.
The applications of Unity in ATLAS~\cite{atlas}, BelleII~\cite{belleII}, and JUNO~\cite{ELAINA} have achieved good visualization effects, and thus Unity is a promising platform for visualization in future HEP experiments.

Despite the advantages of great display effects and cross-platform support, a critical problem in using Unity for detector visualization is that it does not support the commonly used toolkits for HEP detector description, GDML and DD4hep. 
Developers have to use the 3D modeling system in Unity to reconstruct the complicated detector from scratch again, which not only requires extra human work and creates maintenance problems, but could also make the detector description in Unity inconsistent with that in the offline software. 

In this study, we present a method to automate the conversion of the GDML detector description into Unity.
The full detector geometry and its architecture, as described with GDML or ROOT, can be imported into Unity for automatic 3D detector modeling.
The correlation between detector elements and identifiers is also preserved, which makes it convenient for further developments in event displays or virtual reality (VR) applications.
The method was tested and validated using the BESIII detector description~\cite{BESIII_geo,ROOT_Geant4}.

The remainder of this paper is organized as follows.
In Section~\ref{sec:detector}, we introduce the detector description and its visualization in HEP offline software and Unity. 
In Section~\ref{sec:Methodologies}, the method for transformation from GDML to Unity and the detector data flow is presented.
Its application in BESIII is introduced in Section~\ref{sec:BESIII}.
The potential for further development of applications is discussed in Section~\ref{sec:further}.

\section{Detector geometry and visualization}
\label{sec:detector}

\subsection{Detector description in HEP software}

Detector description is an indispensable part of the HEP offline software as it provides the detector geometry and status information for different applications, including simulation, reconstruction, calibration, event display, and physics analysis.
Commonly used HEP infrastructure software products, such as Geant4~\cite{Geant4} or ROOT~\cite{ROOT}, have their own specific detector construction systems.
If software developers define the detector geometry in these types of software independently, potential inconsistencies between them could be created.
To solve this problem, some frameworks provide software-independent detector descriptions, such as GDML and DD4hep, for detector-related applications in the HEP offline software. 

GDML~\cite{GDML} is a detector description language based on extensible markup language (XML)~\cite{XML}.
It describes the detector information through a set of tags and attributes in plain text format to provide a persistent detector description for an experiment.
Several HEP experiments, including BESIII~\cite{BESIII_geo,ROOT_Geant4}, PHENIX~\cite{PHENIX}, LHCb~\cite{lhcb,lhcb_eventdisplay,lhcb_outreach}, and JUNO~\cite{JUNO} , have used GDML to describe and optimize the detector geometry in conceptual design and offline software development ~\cite{rec_qian,rec_event,Rec_impro}.

DD4hep~\cite{dd4hep} is a software framework that provides a complete solution for a full detector description, including geometry, materials, visualization, readout, alignment, and calibration, for the full experimental life cycle. It offers a consistent description through a single source of detector information for simulation, reconstruction, and analysis.
DD4hep is aimed at applications in next-generation HEP experiments, such as CEPC~\cite{CEPC}, ILC~\cite{ILC}, FCC~\cite{FCC}, and STCF~\cite{STCF}.

\subsection{Visualization of detector and events}

One important application of detector description is to visualize it so that users can have a distinct view to better understand HEP detectors.
This is extremely important at the stages of detector conceptual design and commissioning. 
In combination with event visualization, event display provides a powerful tool to demonstrate the detector response to particles, which plays an essential role in offline software development and physics analysis.

However, as part of the traditional industrial design, computer-aided design (CAD)~\cite{CAD} has been dominant in several stages of HEP experiments, including detector design, construction, and commissioning.
There have always been difficulties in sharing 3D modeling information between the CAD and HEP offline software because they belong to two different fields, industrial design and HEP experimentation, respectively.

In HEP experiments, physicists usually develop detector descriptions and event visualization tools in the HEP offline software.
In offline software, developers can make full use of the detector geometry service and event data model to retrieve the corresponding information. 
Event display tools are typically based on commonly used HEP software, Geant4 or ROOT, whose visualization functions are friendly to HEP users and are naturally convenient for visualization software development. 
For example, the BESIII event display software is based on the infrastructure visualization function of ROOT.
With the upgrade of ROOT and its EVE package~\cite{ROOT_EVE}, the development of event display tools has become more convenient.
Several recent HEP experiments, such as ALICE~\cite{ALICE}, CMS~\cite{CMS}, JUNO~\cite{JUNO}, and Mu2e~\cite{Mu2e}, have developed event-display software based on ROOT EVE.

However, owing to the limited visualization support of ROOT, its display effects are unsatisfactory for meeting the diverse requirements of physicists.
Most ROOT applications are limited to Linux platforms.
For better visualization and interactivity, several event display tools have been developed based on an external visualization software.
Atlantis~\cite{Atlantis,Atlantis_event} and VP1~\cite{VP1} are general-purpose event display tools in ATLAS.
CMS has also developed several visualization systems such as Fireworks~\cite{Fireworks} and SketchUp~\cite{SketchUpCMS}.

\subsection{3D modeling and visualization in Unity}
\label{sec:3DUnity}

\begin{figure}[!htb]
	\includegraphics[width=0.9\hsize]{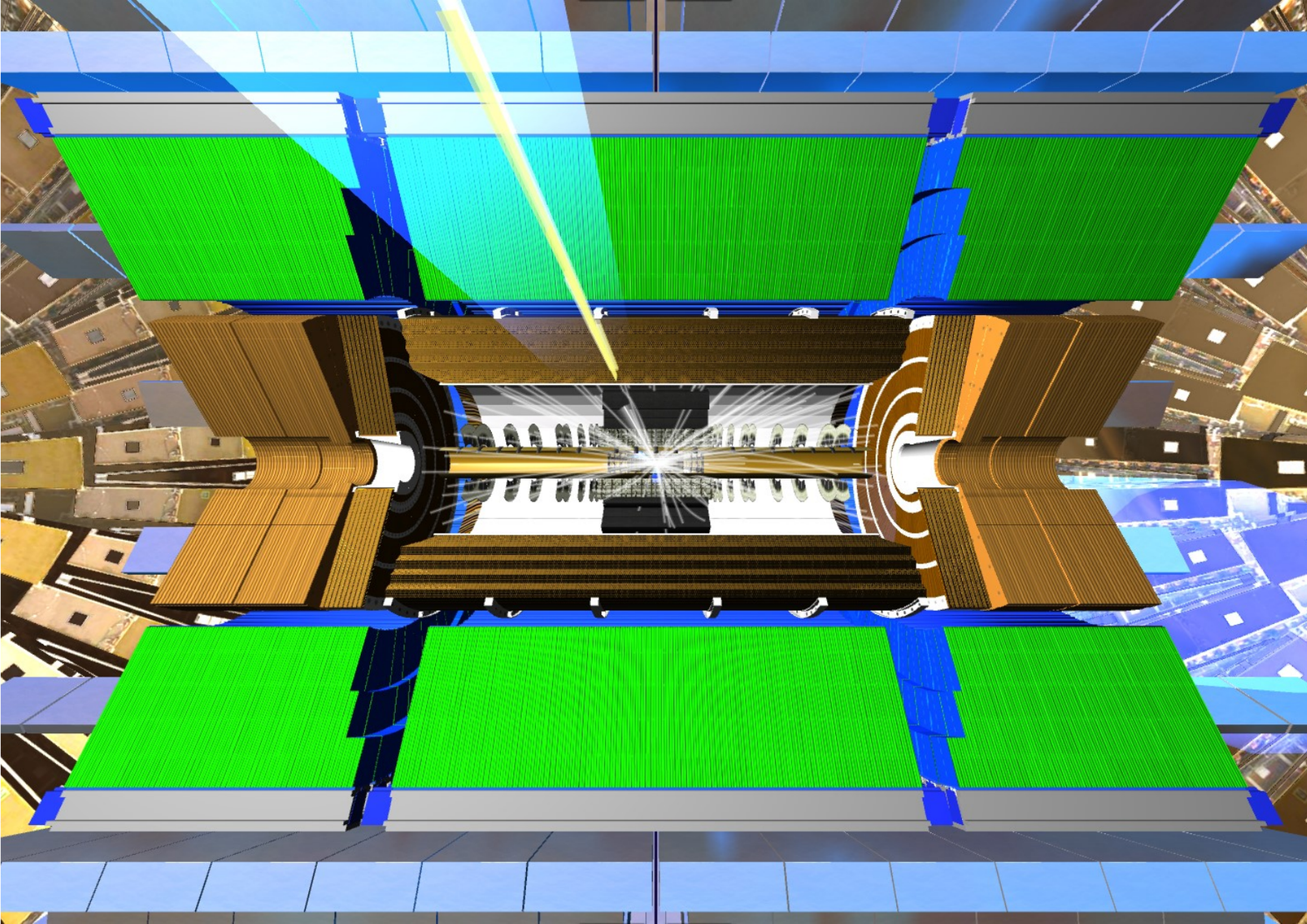}
	\caption{CAMELIA~\cite{CAMELIA}, the ATLAS event display tool based on Unity.}
	\label{fig:camelia}
\end{figure}

\begin{figure}[!htb]
	\includegraphics[width=0.9\hsize]{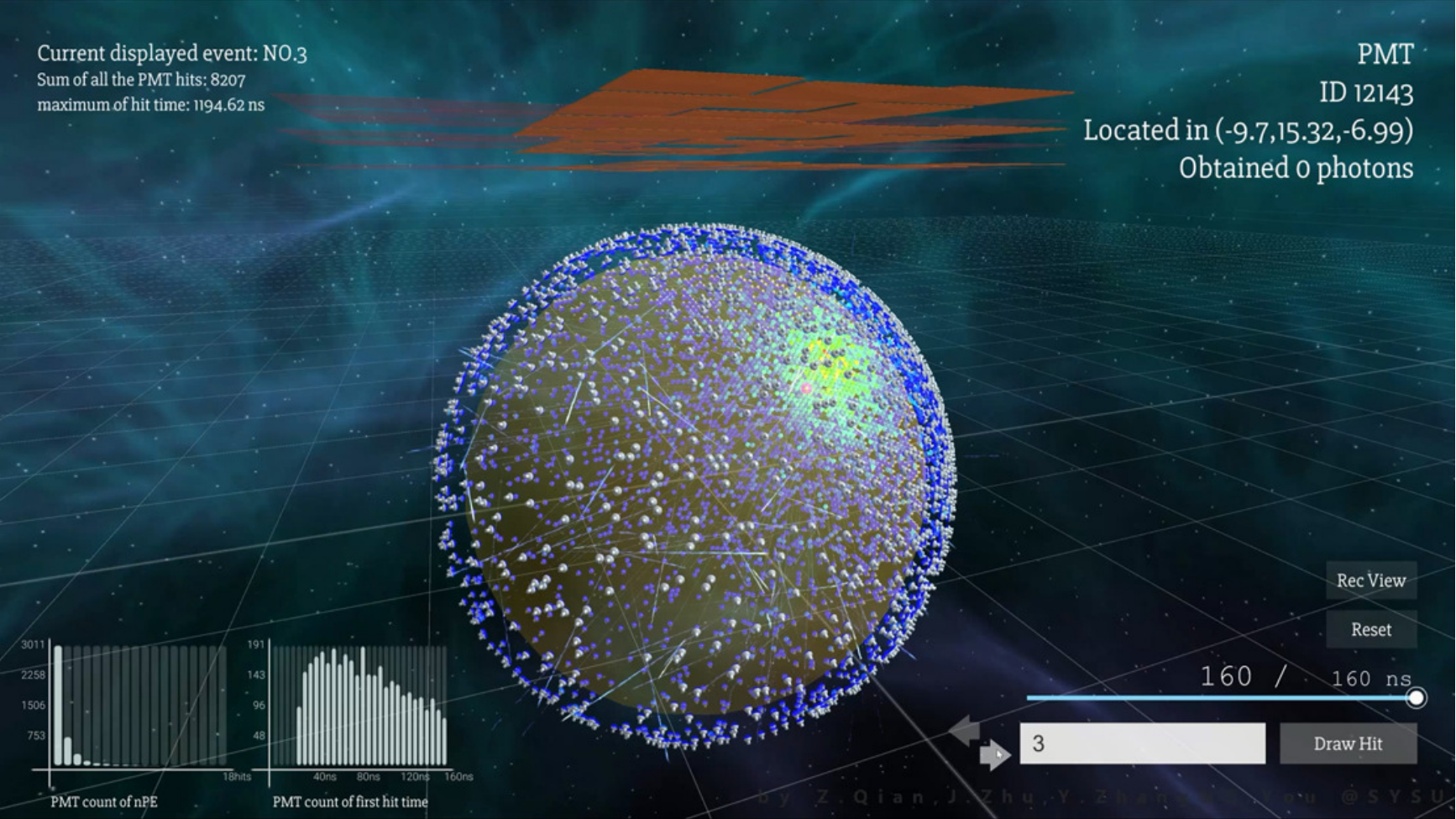}
	\caption{ELAINA~\cite{ELAINA}, the JUNO event display tool based on Unity.}
	\label{fig:elaina}
\end{figure}

Unity is a professional video and game production engine that supports more than 20 platforms.
It is particularly popular for iOS and Android mobile app development and has recently been used in HEP for scientific research and outreach.
The Cross-platform Atlas Multimedia Educational Lab for Interactive Analysis (CAMELIA) ~\cite{CAMELIA,CAMELIA_web} is an event display software based on Unity for ATLAS.
Fig.~\ref{fig:camelia} shows the visualization of the ATLAS detector and a proton--proton  collision event in CAMELIA.
Another event display tool based on Unity, the Event Live Animation with Unity for Neutrino Analysis (ELAINA)~\cite{ELAINA} has also been developed in JUNO, as shown in Fig.~\ref{fig:elaina}.

The visualization software based on Unity has several advantages. 
\begin{itemize}

\item \emph{Impressive visualization effects.} The Unity engine, as a professional 3D software, provides a more detailed description of objects and striking visualization effects, which makes it much more powerful than the traditional event display based on ROOT.

\item \emph{Cross-platform support.} Thanks to the multi-platform support of Unity, after the development by building models and functions in Unity, a project can be directly exported and deployed in different operating systems, including Windows, Linux, macOS, iOS, Android, and network browsers.
This feature makes the same visualization project available for all the users on different platforms.
This not only reduces the workload in project development, but also facilitates the maintenance.

\item \emph{Extensibility.} Unity allows the project to be extended and further developed into VR~\cite{VR} or augmented reality (AR)~\cite{AR} projects, such as BelleII VR~\cite{BelleIIVR}, thus providing a quite different way for detector design, offline software development, and physics analysis.
It can also be very helpful for advertising scientific projects to the public, especially the physics behind the large-scale scientific apparatus. It could be a powerful tool in education and outreach, such as the Total Event Visualiser (TEV) of the CERN Media Lab~\cite{TEV}. 
Support for various VR or AR devices has been integrated in Unity, and they will be continuously supported with Unity upgrades, so that the visualization project developers can focus on the software project itself, instead of the support for different hardware.  

\end{itemize}

Because of its prominent features, Unity has huge potential and boosts promising prospects in the development of visualization software.
Not only can it be used for scientific research in particle and nuclear physics, but it also has wide application potential in nuclear techniques for the industry.

\section{Methodologies}
\label{sec:Methodologies}

Although Unity has great advantages for the development of visualization software, many endeavors have to be made for its application to large-scale scientific apparatuses.
For example, HEP detectors are typically highly complicated, with millions of components, which makes it difficult to build a 3D model of the detector in Unity.

To develop the event display tool for HEP experiments, the detector description in Unity should be fully consistent with that in the offline software, so that the displays of the tracks, hits, showers, and detector components match each other.   
Usually, for an HEP experiment, the detector description already exists in offline software. Therefore, we propose a method to convert the HEP detector description into 3D modeling in Unity automatically.

\subsection{Automatic detector geometry transformation}

In HEP offline software, to avoid inconsistency of the detector description in data processing, it is essential that the framework can provide a single source of detector description for all applications.
This idea is typically realized by developing automatic geometry transformation interfaces between different software packages.  
For an existing detector description with GDML, the detector construction in Geant4 and ROOT for a specific experiment can be automatically completed with the GDML--Geant4 and GDML--ROOT interfaces~\cite{BESIII_geo,ROOT_Geant4}, respectively.
In this way, detector geometry consistency between  the Geant4 based simulation, ROOT-based reconstruction, event display, and data analysis is automatically guaranteed. 

A similar idea can also be implemented in the 3D detector modeling of Unity.  
HEP experiments usually already have a detailed detector description in some formats, such as GDML.
If a novel method for GDML--Unity conversion can be realized, and given that it is a universal technique, all the current and future HEP experiments can benefit, making it possible the easy development of applications for detector visualization, event display, and outreach.

\subsection{Detector data conversion from GDML to Unity}

In the industry market, there are tens of popular 3D file formats; however, none of them are commonly supported in both HEP software and Unity.
Therefore, we must find a data flow path that starts from GDML and ends in Unity with minimum steps of conversion.

FreeCAD~\cite{FreeCAD} is a general-purpose parametric 3D CAD  and modeling software. 
It is free and open-source, and users can extend its functionality.
Because FreeCAD supports the import of geometry in constructive solid geometry (CSG) format~\cite{CSG}, an interface between GDML and FreeCAD was developed by Keith Sloan et al.~\cite{GDML-FreeCAD} to import the GDML-format detector description.
Meanwhile, FreeCAD allows to export the modeling in several types of widespread 3D formats, including  STEP~\cite{STEP}, BREP~\cite{BREP}, and VRML~\cite{VRML}.

Among these formats, STEP was adopted for further geometry transformation.
Although STEP is still not a format that can be directly read by Unity, it can be transformed into the Filmbox (FBX) format~\cite{FBX} with Pixyz Studio software~\cite{pixyz}.
Pixyz Studio is a CAD data preparation and optimization software that supports more than 35 3D file formats.
It provides tesselation algorithms, enabling the transformation of CAD data from industry-leading solutions into lightweight, optimized meshes such as FBX, which is a 3D mesh format supported by Unity.
It is noteworthy that Pixyz joined Unity so that Unity can provide plugins to import and transform heavy and complex 3D CAD data, such as HEP detectors, into optimized meshes for real-time 3D engines. 

\begin{figure}[!htb]
	\includegraphics[width=0.98\hsize]{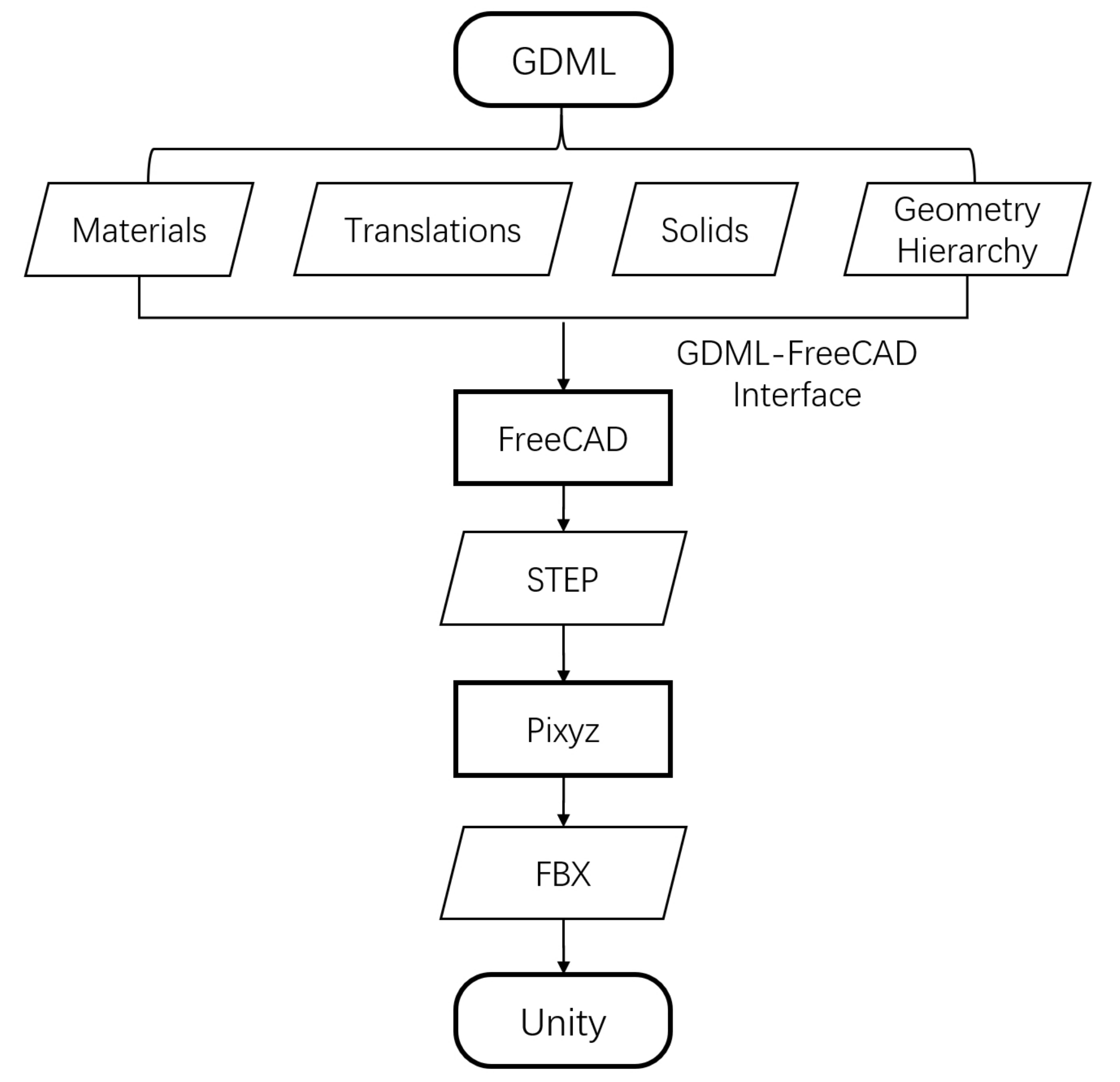}
	\caption{Detector data flow from GDML to Unity.}
	\label{fig:meshing}
\end{figure}

The full chain of data flow is shown in Fig.~\ref{fig:meshing}.
The original HEP detector description in the GDML format usually consists of the following parts: the material list,  position and rotation list, shape list, detector component (physical node) list, and hierarchy of the whole detector tree. 
The GDML detector description was first imported into FreeCAD using the GDML--FreeCAD interface and then  exported into the STEP format.
Pixzy reads in the STEP format data and transforms it into the FBX format, which can be read directly by Unity retaining the detector unit association information.   

Using the data flow chain described above, we provide a method to transform the GDML detector description into Unity with automated 3D detector construction. 
The method was realized using the GDML--FreeCAD interface, FreeCAD, and Pixyz software.
The correctness of the detector geometry conversion in each step was validated by visualization and comparison at the shape  and detector levels.   
With the 3D detector modeling successfully constructed in Unity with automation, further application development based on Unity is achievable. 

\subsection{Integrity in Unity 3D modeling}

For a specific HEP detector with a GDML description, to realize a complete 3D model of the detector in Unity, additional work is necessary in addition to the automatic geometry transformation method introduced above.

First, in the GDML--FreeCAD interface~\cite{GDML-FreeCAD}, not all solid shapes that are currently defined in Geant4 or ROOT are fully supported.
The current interface only supports approximately 10 types of shapes, which are mostly basic CSG shapes, such as boxes, cones, cylinders, ellipsoids, and tubes.
More than 30 types of specific shapes are provided in Geant4 but are not commonly used.
For a specific HEP detector description with the GDML format, if some of the special shapes used are not available in the interface, users may request to develop the corresponding shape transformation in the GDML--FreeCAD interface, so that the full detector transformation can be supported.

\begin{figure}[!htb]
	\includegraphics[width=0.99\hsize]{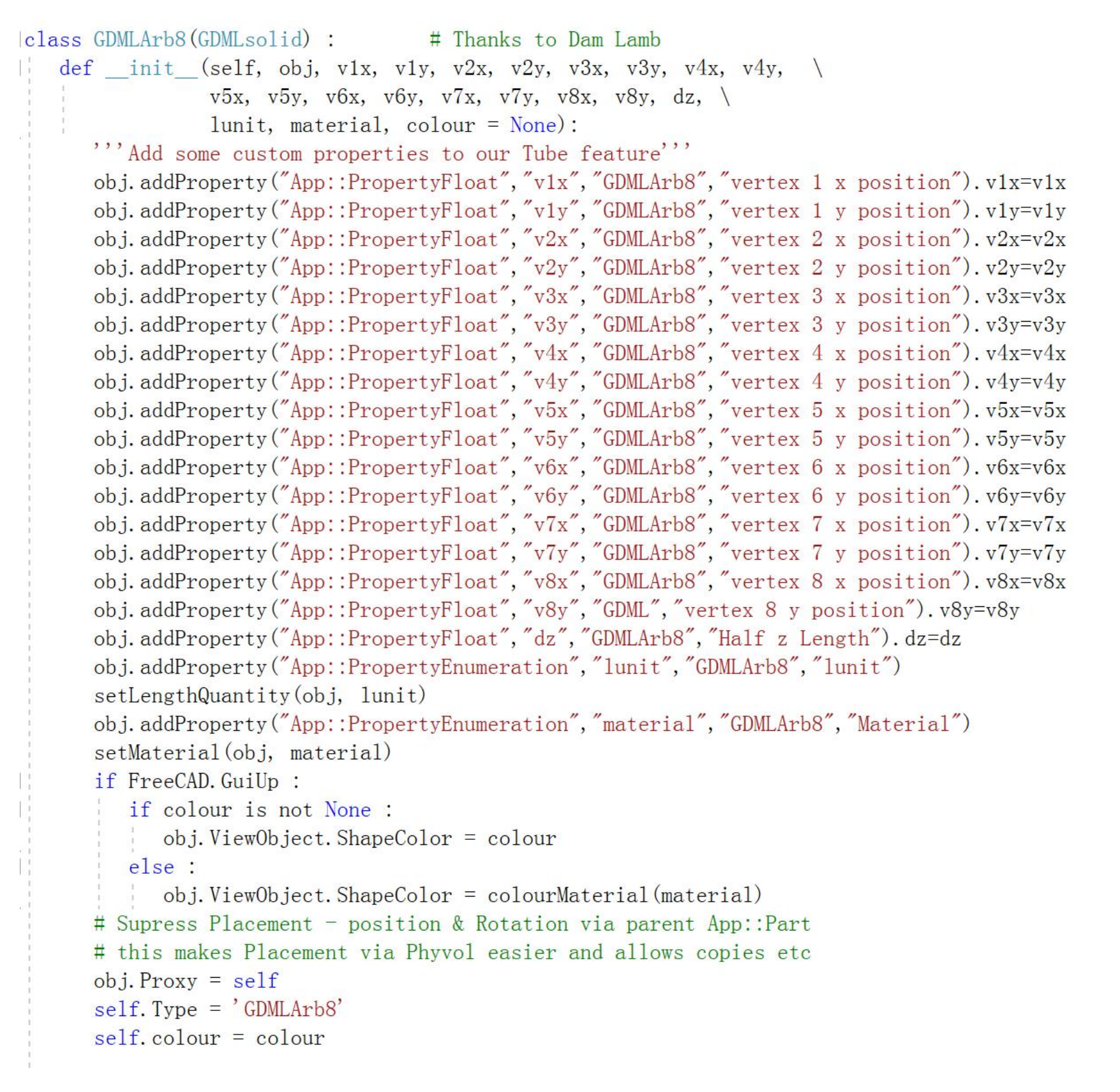}
	\caption{Code example for conversion of the Arb8 shape in the GDML--FreeCAD interface~\cite{GDML-FreeCAD}.}
	\label{fig:arb8code}
\end{figure}

\begin{figure}[htbp]
  \centering
\includegraphics[width=0.50\columnwidth]{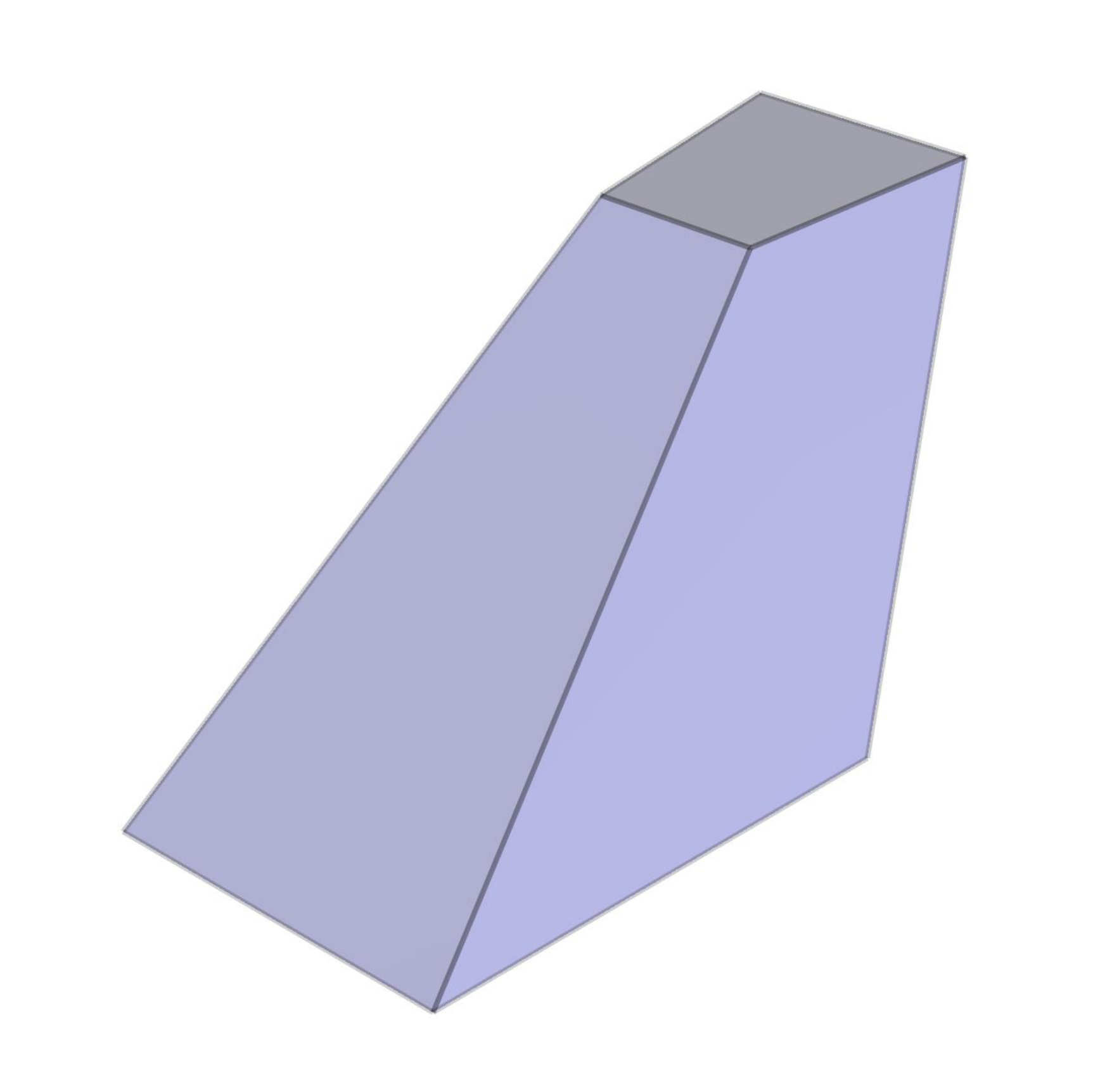}
\includegraphics[width=0.42\columnwidth]{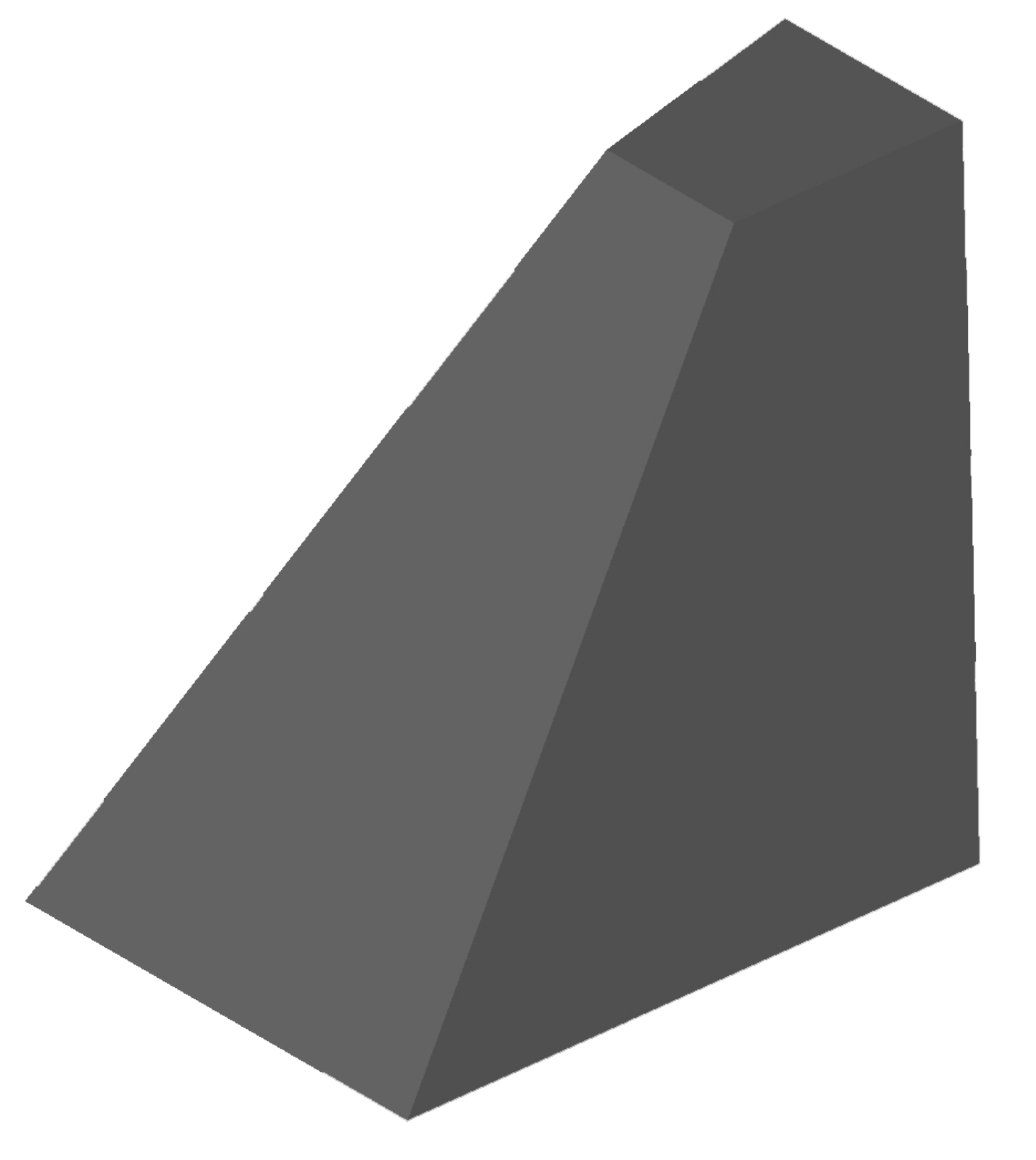}
  \caption{\small
Visualization of the Arb8 shape in ROOT (left) and FreeCAD (right).
}
\label{fig:arb8}
\end{figure}

For example, Arb8 is an arbitrary trapezoid defined by eight vertices standing on two parallel planes perpendicular to the Z-axis.
The Arb8 shape exists in GDML, ROOT, and FreeCAD. 
In the GDML--FreeCAD interface, the conversion interface is updated with Arb8 shape support, and the code example is shown in Fig. ~\ref{fig:arb8code}).
The same Arb8 shape with the GDML description and its visualization in ROOT and FreeCAD are also compared in Fig. ~\ref{fig:arb8}. 

Second, in the description of an HEP detector, it is important to maintain the association between the detector unit and its unique identifier in the event data model.
Such association information is essential for controlling the visualization properties of the detector unit in further application development in Unity, such as event displays.
Usually, identifier information is stored in the name of each node in GDML.
As long as the node name remains unchanged during the entire chain of conversion, the identifier can be extracted later in Unity from the unit's name to retrieve the mapping of the unit and its corresponding identifier in the offline software. This makes it possible to control the visual effects of a geometry unit with its identifier in event data. 
Hence, in the transformation of the detector geometry from GDML to Unity, it is critical to retain the name of each detector unit to retain association information.

Third, although basic information such as density and makeup of materials can be transformed from GDML, Unity provides richer visualization properties such as material color, texture, transparency, and reflection, which allow users to have the freedom to render the detector geometry with more powerful visualization effects. 

With the above supplementary information provided, 3D detector modeling has been successfully constructed in Unity and   has been qualified for further development.

\section{Application in BESIII}
\label{sec:BESIII}
 
\subsection{BESIII detector description and visualization}

BESIII is a spectrometer operating at the Beijing Electron Positron Collider II (BEPCII). 
The BESIII detector records symmetric $e^+e^-$ collisions provided by the BEPCII storage ring~\cite{bepcii}, which operates in the center-of-mass energy range of 2.0 to 4.7~GeV~\cite{BESIII_future}.
The cylindrical core of the BESIII detector covers 93\% of the full solid angle and consists of a helium-based multilayer drift chamber~(MDC), a plastic scintillator time-of-flight system~(TOF), and a CsI(Tl) electromagnetic calorimeter~(EMC), which are all enclosed in a superconducting solenoidal magnet providing a 1.0~T magnetic field. 
The solenoid is supported by an octagonal flux-return yoke with resistive plate counter muon identification modules interleaved with steel~(MUC). 

The detector description in the BESIII offline software is provided in GDML format.
Each of the four subdetectors, MDC, TOF, EMC, and MUC, has a corresponding GDML file to describe it.
A general GDML file provides a description of common materials and other passive detector components, such as the beam pipe and superconducting solenoid.
All applications in BESIII offline software, including simulation, reconstruction, event display, calibration, and analysis, use GDML files as the single source of detector information.

\begin{figure}[!htb]
	\includegraphics[width=0.9\hsize]{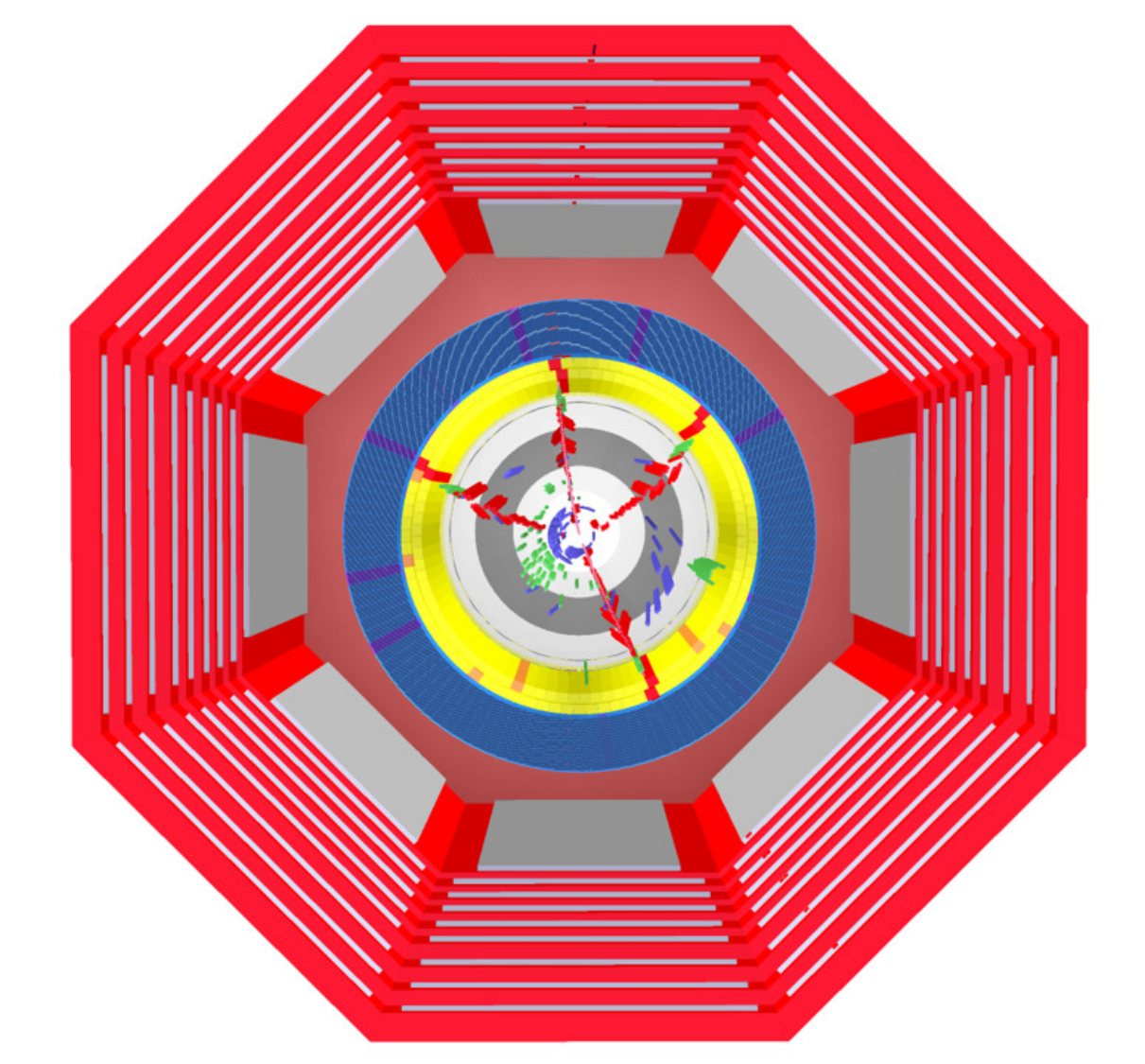}
	\caption{Visualization of the BESIII detector and a real data event in BesVis.}
	\label{fig:besvis}
\end{figure}

BESIII Visualization (BesVis), an event display tool based on ROOT, was developed to visualize the detector and analyze physics events, as shown in Fig. ~\ref{fig:besvis}. 
It has played an important role in BESIII’s offline software development and physics analysis since 2005.

\subsection{Conversion of the BESIII geometry to Unity}

Based on the method introduced by Section~\ref{sec:Methodologies}, it is possible to perform a format conversion of the BESIII detectors description from GDML to Unity.
The detector data conversion process comprised the following steps:

First, the GDML description of each subdetector was imported into FreeCAD with the GDML--FreeCAD interface.
Because the original interface only supports the basic CSG shapes, BESIII has used some complicated shapes, such as Twistedtubs, IrregBox, and Boolean shapes. We updated the interface to allow correct conversion of all the shapes being used in the GDML detector description of BESIII.

\begin{figure}[htbp]
  \centering
\includegraphics[width=0.45\columnwidth]{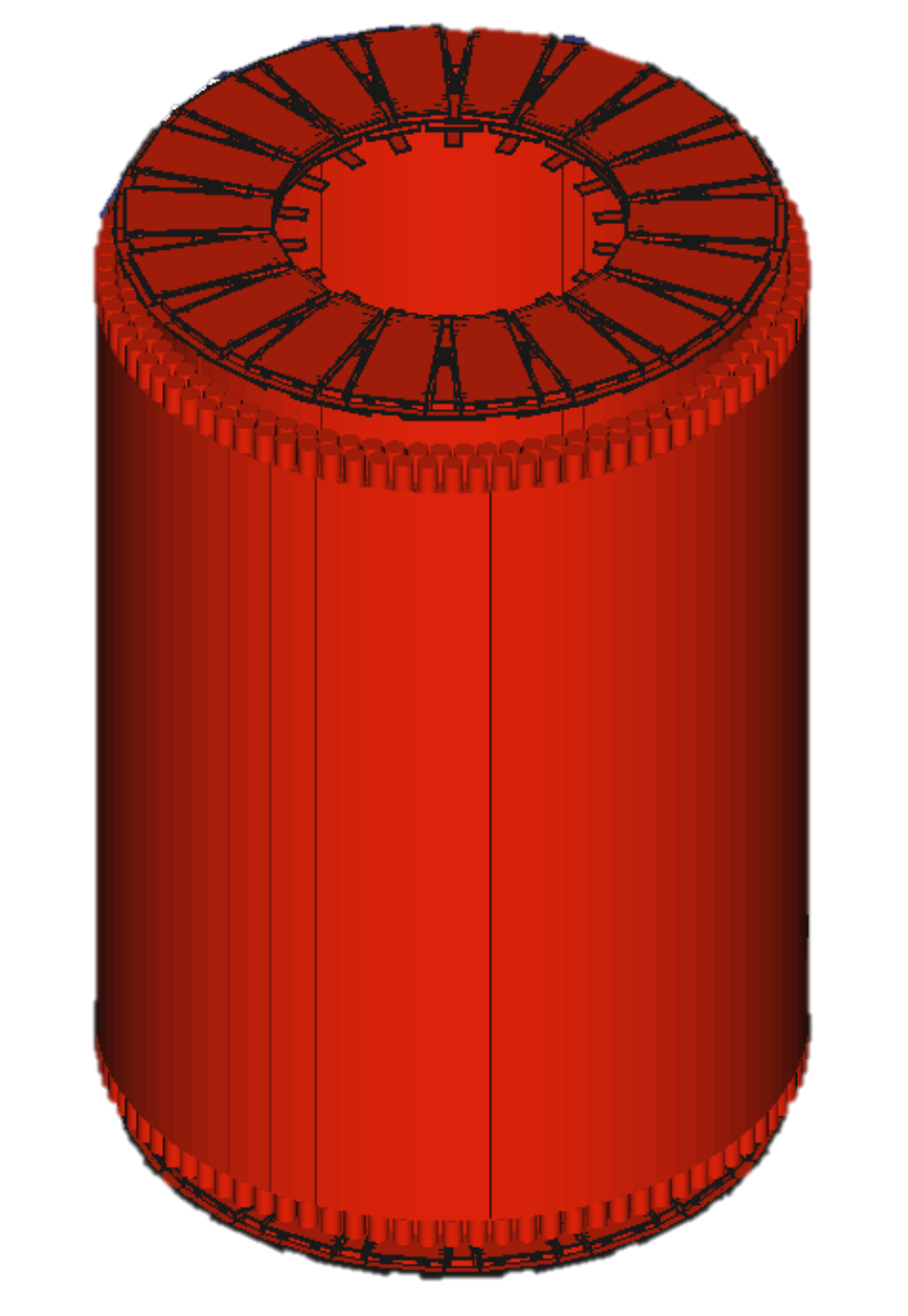}
\includegraphics[width=0.43\columnwidth]{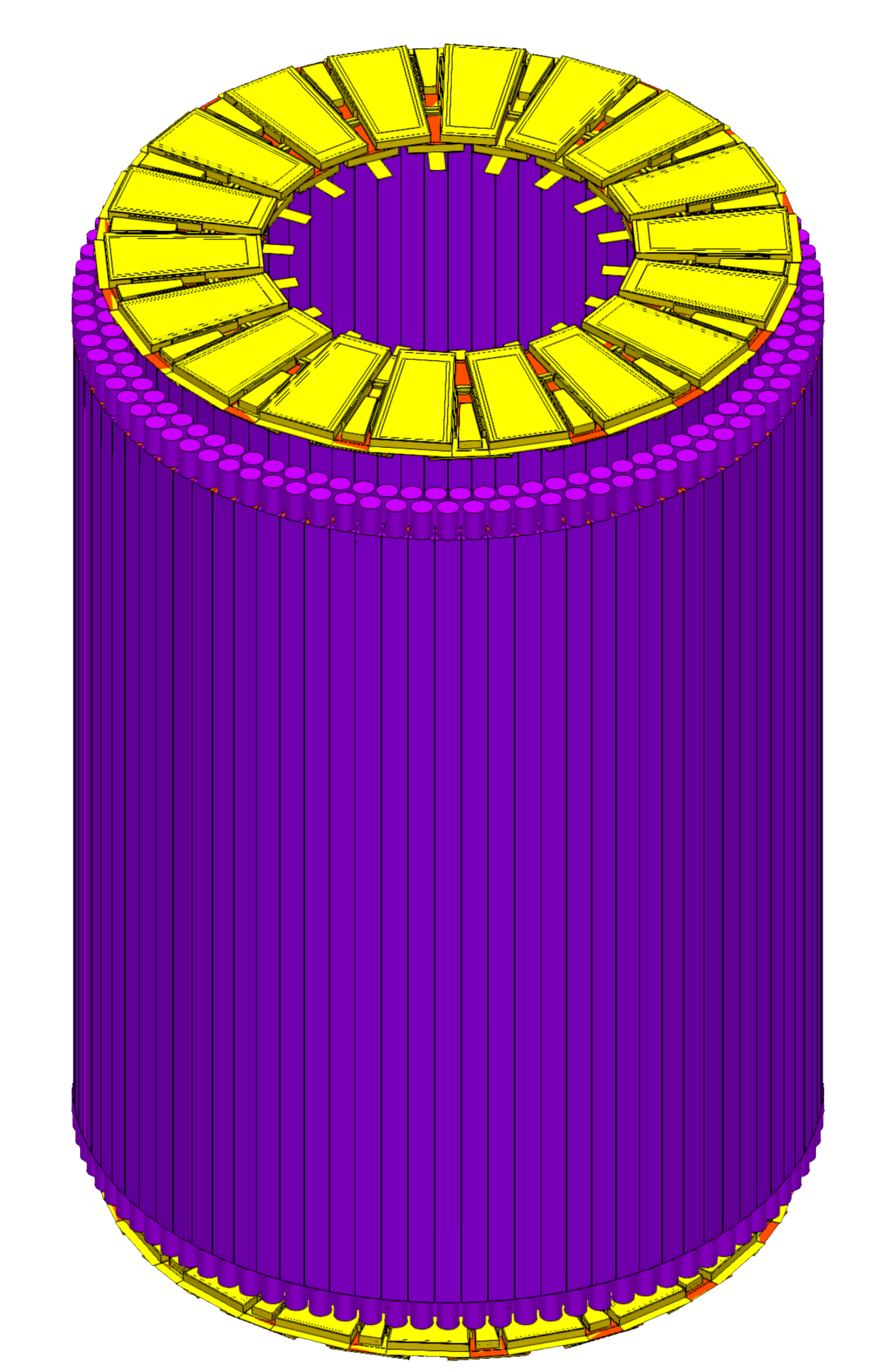}
  \caption{\small
Visualization of the BESIII TOF sub-detector in FreeCAD (left) and Pixyz (right).
}
\label{fig:TofFreeCADPixyz}
\end{figure}

Second, after importing into FreeCAD, the BESIII detector data were exported in the STEP format.
Then, the STEP files were converted to FBX files using the format conversion function provided by Pixyz Studio.
As an example, the visualization of the BESIII TOF subdetector in FreeCAD and Pixyz is shown in Fig. ~\ref{fig:TofFreeCADPixyz}.

Third, the FBX files were directly imported into Unity.
For the BESIII detector, four sub-detector FBX files (MDC, TOF, EMC, and MUC) and one general FBX file describe the beam pipe and superconducting solenoid.
The subdetectors were verified in Geant4 and ROOT to guarantee no space overlaps between each other, so the subdetectors could be directly combined to form the whole BESIII detector.
Finally, 3D modeling  of the BESIII detector was successfully performed in Unity after a set of automated steps. 

\subsection{Display of the BESIII detector in Unity}

Although at this stage the BESIII detector has been automatically built in Unity, it is still not well displayed in the scene.
Because GDML does not save visualization attributes, all the detector units are still set with the default visualization attributes in Unity.
The only object that users can see in the scene is a box, which is the top world volume in the definition of the GDML detector description.

A set of scripts must be developed to set the color, transparency, reflectivity, and texture of the material for the detector units. 
The top world volume and virtual mother volumes need to be set invisible to allow the inner detector units to appear. 
Another advantage of automatic conversion is that the name of each detector unit is maintained during the process of transformation so that the scripts can set the visualization attributes according to the name of each detector unit. 

\begin{figure}[htbp]
  \centering
\includegraphics[width=0.50\columnwidth]{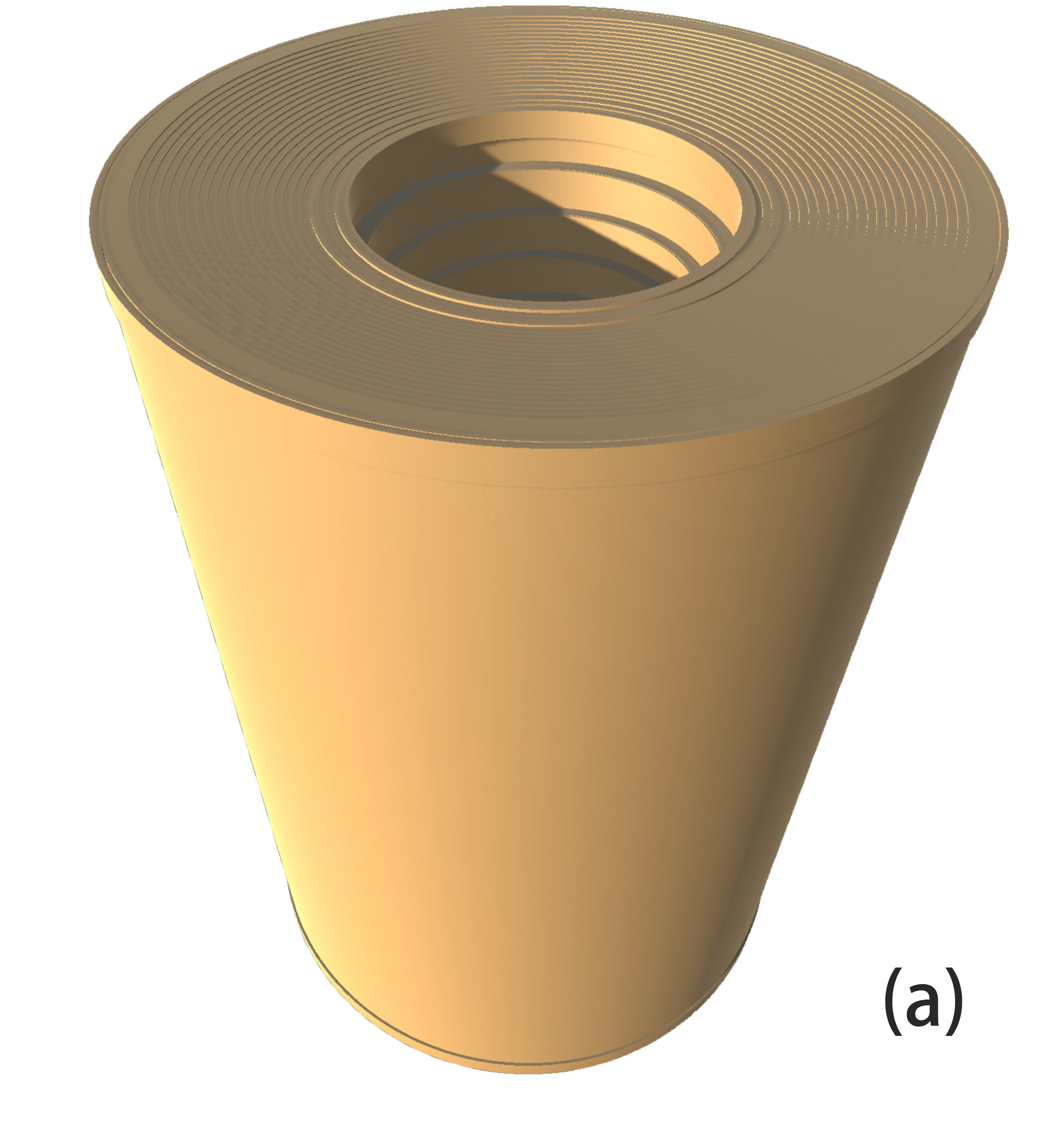}
\includegraphics[width=0.47\columnwidth]{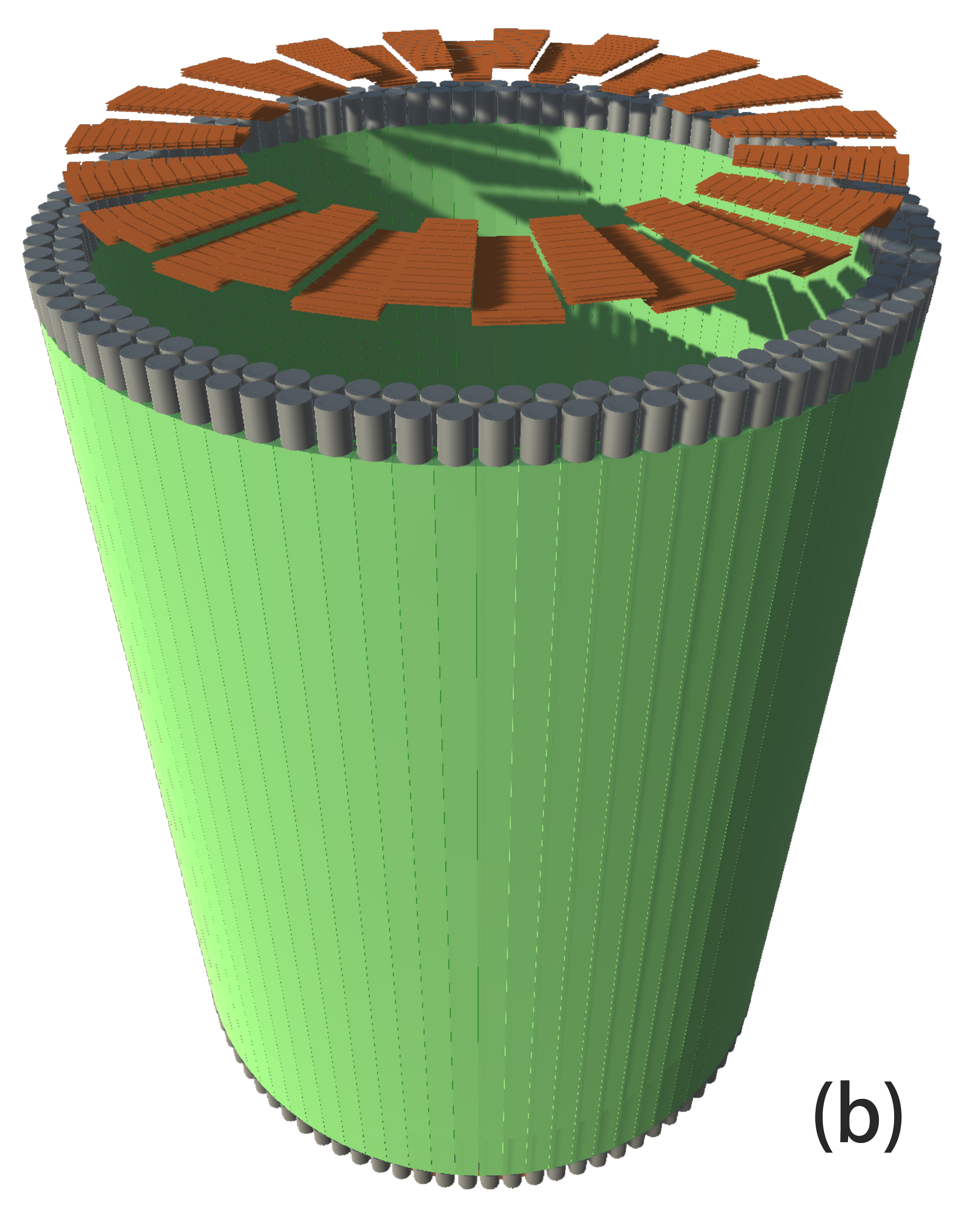}
\includegraphics[width=0.45\columnwidth]{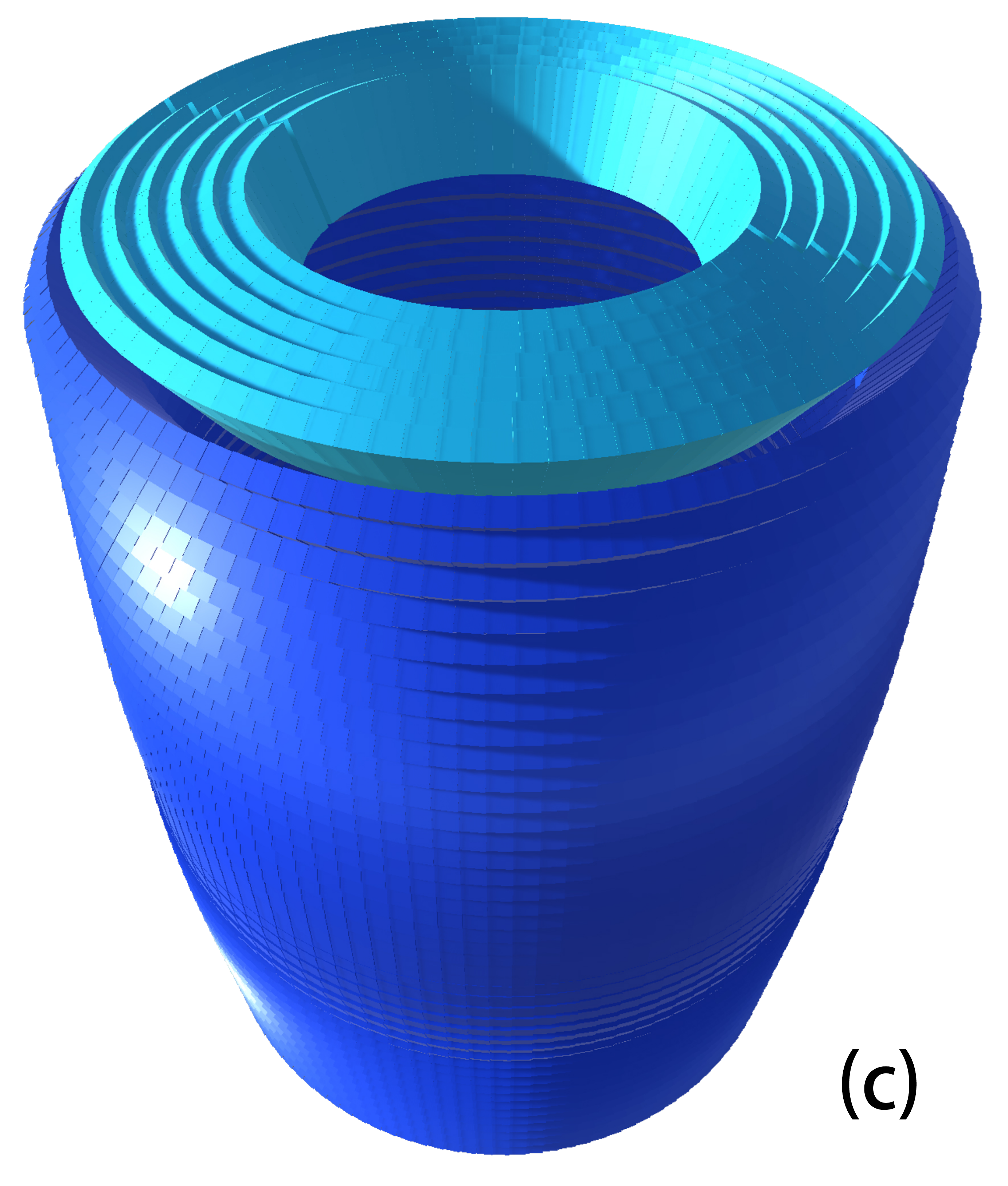}
\includegraphics[width=0.53\columnwidth]{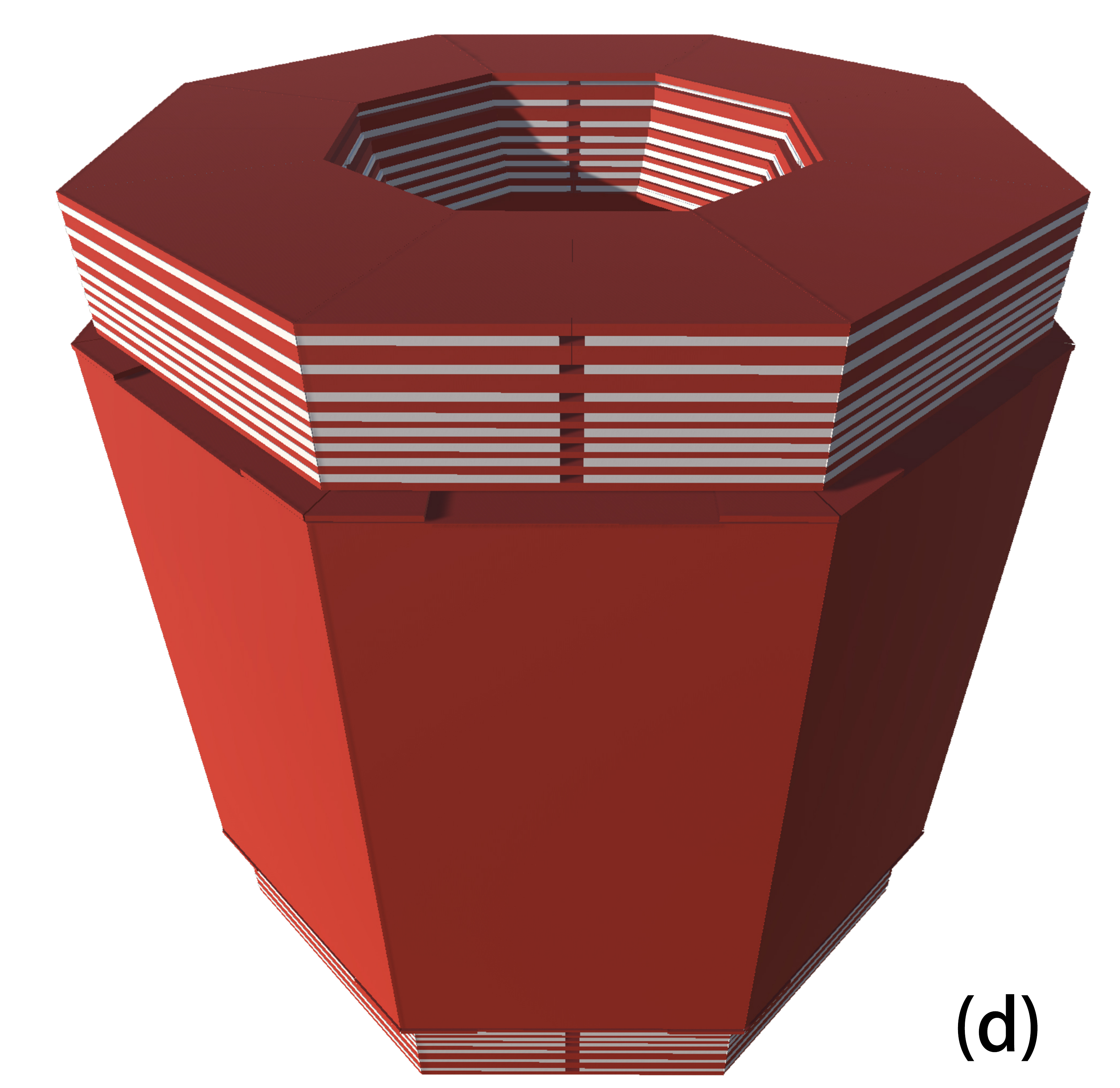}
  \caption{\small
Visualization of the BESIII sub-detectors in Unity, (a) MDC, (b) TOF, (c) EMC, and (d) MUC. 
}
\label{fig:SubDetectorUnity}
\end{figure}

\begin{figure}[!htb]
	\includegraphics[width=0.9\hsize]{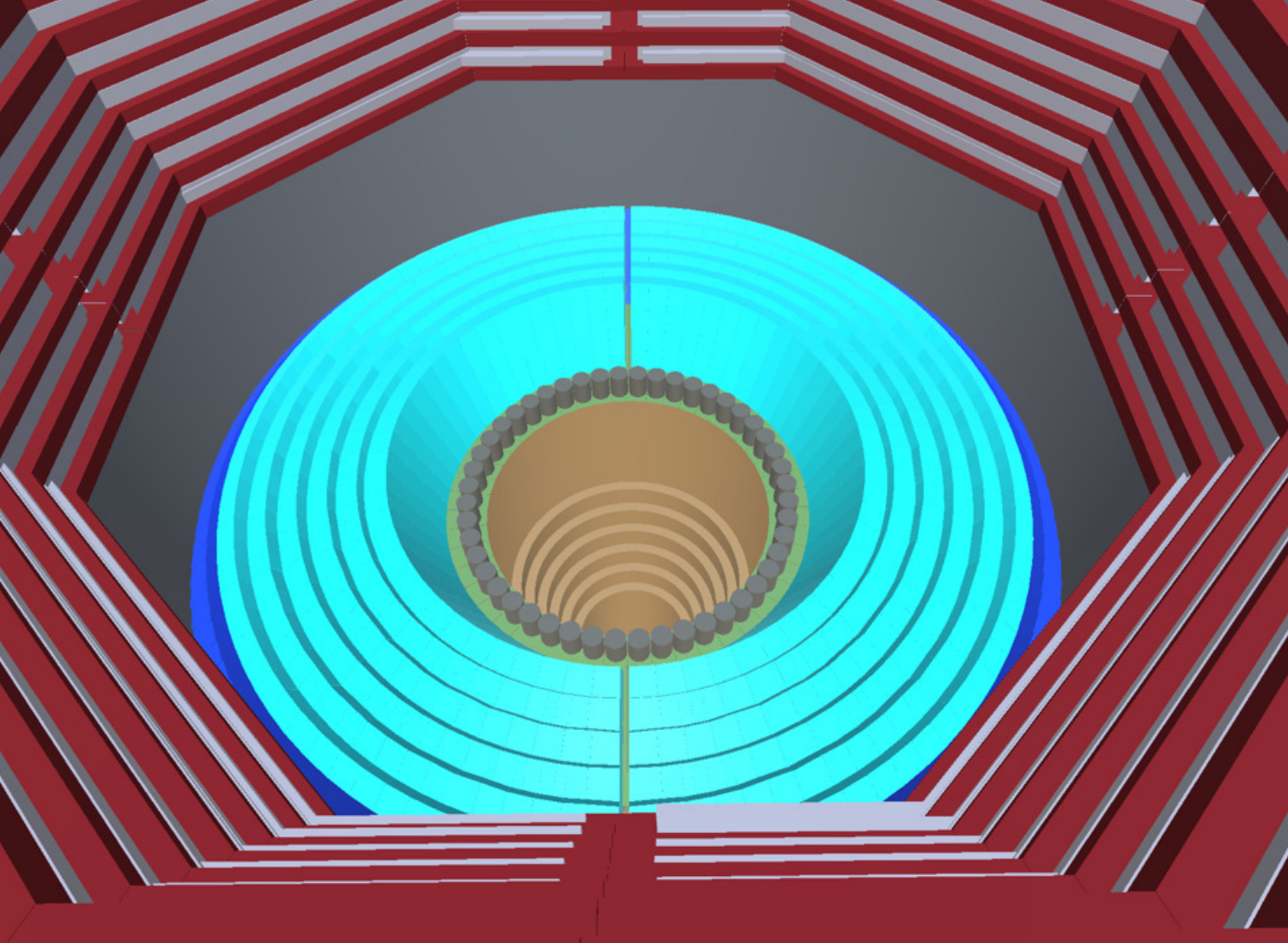}
	\caption{
Display of the full BESIII detector in Unity.
From inside to outside: MDC, TOF, EMC and MUC.
}
	\label{fig:BESinUnity}
\end{figure}

Fig.~\ref{fig:SubDetectorUnity} shows the display of the four BESIII subdetectors in Unity. 
The corresponding display parameters of the materials were updated to obtain a better display effect.
The display attributes in Unity were set to be consistent with those in BesVis, but with more vivid and richer visualization effects than those supported by ROOT.
The full BESIII detector in Unity is shown in Fig. ~\ref{fig:BESinUnity}.

\section{Further development of applications}
\label{sec:further}

Once the detector geometry and its visualization attributes are constructed in Unity, various applications can be further developed.
The characteristics of Unity, including fantastic visual effects, multi-platform support, and VR~\cite{VR} or AR~\cite{AR} device integration, make it convenient for developing event display tools, detector status monitoring software, and VR/AR applications for scientific research and education.

\subsection{Event display tool}

Event displays are convenient tools for offline software tuning and physical analysis in HEP experiments.
In Section~\ref{sec:3DUnity}, the application of Unity for event displays in ATLAS and JUNO was introduced.
However, for both programs, the detectors were manually constructed independently in Unity, which required a large amount of  developers' work.

Using the automatic geometry transformation method, the BESIII detector was constructed in Unity.
Compared with the geometry construction in the BESIII offline software and its ROOT-based event display, which is composed of more than 5000 lines of code, this method avoids the repetitive coding work required in a different software.
To develop an event display tool based on Unity, the event data information is obtained and the fired detector units are associated with their identifiers, which can be decoded from the name of each detector unit.

\begin{figure}[!htb]
	\includegraphics[width=0.9\hsize]{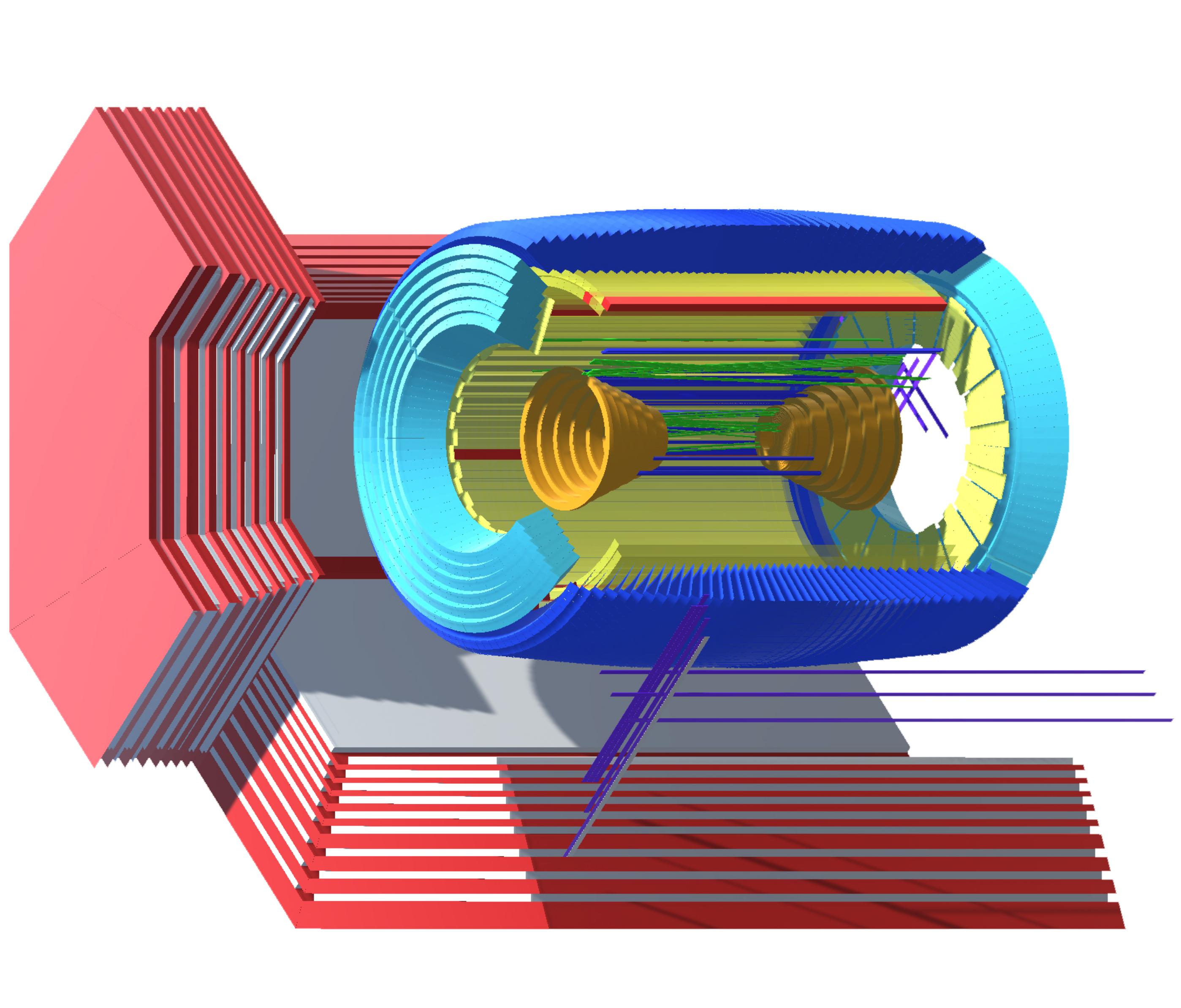}
	\caption{Prototype of a new BESIII event display tool in Unity and its rendering effects.}
	\label{fig:BESUnityEventDisplay}
\end{figure}

Once the association relationship is constructed, a set of scripts can be developed in Unity to control the different visual effects of the fired and unfired detector units.
The basic functions of the event display tool can then be realized.
Fig.~\ref{fig:BESUnityEventDisplay} shows a prototype of the BESII event display tool under development in Unity, and its rendering effects.   

\subsection{Detector and data monitoring}

With the detector constructed in Unity, not only the offline event display but also the online monitoring software can be developed to help monitor the operation status of the experiment and the quality of the real-time data collected by the detector.

Detector units with abnormal operational statuses, such as dead or hot channels, can be distinctly displayed, which will help shift operators diagnose potential detector problems.

Owing to the multi-platform support of Unity, a monitoring project can be easily deployed on different platforms and devices.
In addition to the traditional Windows and Linux operating systems, the monitoring project can also be built into apps on mobile platforms, such as Android and iOS, so that users can conveniently monitor the status of an experiment remotely from their mobile phones or pads.

\subsection{Virtual reality applications}

In the roadmap for HEP software and computing R\&D for the 2020s~\cite{Roadmap}, an application based on VR is an important development direction ~\cite{HSF}.
VR simulates the physical presence of a user in a virtual environment.
By presenting the emitting particles and their interactions with the detector, it can provide a new method with immersive experience for physicists to tune the offline simulation and reconstruction software. They can also perform physics analysis for rare events, as if the users were personally in the scene of the running detector. 
Some creative productions have started in the HEP community, such as ATLASrift~\cite{ATLASrift} and BelleII VR~\cite{BelleIIVR}.

With the method for automatic detector construction in Unity and the extensible support of VR devices from Unity, VR applications for nuclear and HEP experiments can also be conveniently developed.
Because most of the nuclear or HEP experiments are not accessible during data collection  because of safety or security requirements, our method has another advantage, allowing the general public to explore the detector with an immersive experience and watch nuclear or HEP collisions in a realistic environment.
It will be of great benefit to let the general public have a basic idea and understanding of the type of scientific research that the nuclear physics and HEP communities are doing.

\subsection{Interdisciplinary applications}

In addition to its applications in nuclear physics or HEP experiments, this method could also have wide applications in applied nuclear science, techniques, and industry.

For example, in nuclear power plant (NPP) monitoring, this method can be used to monitor the emitting neutrons or neutrinos and then reconstruct the distribution of the inner active fission area to monitor the operation status of the NPP.

In the field of nuclear medical imaging, the human body is more or less like a nuclear particle source or detector, but with dynamic geometry.
Because our detector modeling method is automatic, it can rapidly construct and visualize the human body, allowing observation of its status after interaction with nuclear particles in real time or semi-real time.  
Other potential interdisciplinary applications include muon tomography, X-ray inspection, and fusion diagnosis~\cite{MtEtna_muons, Khufu_muons, laoheishan_muons, juno_muon_bundle}. 

\section{Summary}

Detector description and visualization are essential techniques in software development for next-generation HEP experiments.
Unity is a popular and versatile multimedia creation platform that provides impressive visual effects, multiple platform support, and extensibility to VR and AR applications.
For large-scale scientific projects, building a complicated detector with up to millions of components in Unity while keeping them consistent with the geometry in the offline data processing software is very difficult and requires significant work for software developers.

A method for automatic detector data transformation from GDML to Unity is presented, which can construct 3D modeling of detectors in Unity for further applications.  
The method has been realized in the BESIII detector with hundreds of thousands of components and can be used in future experiments such as CEPC.
In addition to HEP experiments, the method also has great potential for multiple interdisciplinary applications, education, and outreach.

\section*{Acknowledgements}

This work was supported by the National Natural Science Foundation of China (Grant Nos. 11975021, 12175321, 11675275, U1832204, and U1932101), National Key Research and Development Program of China (Nos. 2020YFA0406300 and 2020YFA0406400), Guangdong Basic and Applied Basic Research Foundation (2021A1515012039), State Key Laboratory of Nuclear Physics and Technology, Peking University  (Nos. NPT2020KFY04 and NPT2020KFY05), Strategic Priority Research Program of the Chinese Academy of Sciences (No. XDA10010900), 
Chinese Academy of Sciences (CAS) Large-Scale Scientific Facility Program; Fundamental Research Funds for the Central Universities, Sun Yat-sen University, National College Students Science and Technology Innovation Project, and Undergraduate Base Scientific Research Project of Sun Yat-sen University.

\printbibheading[heading=bibintoc]
\printbibliography[heading=none]

\end{document}